\documentclass[aps,prl,twocolumn,superscriptaddress]{revtex4-2}
\usepackage[T1]{fontenc}
\usepackage{graphicx}
\usepackage{dcolumn}
\usepackage{bm}
\usepackage[utf8]{inputenc}
\usepackage{upgreek}
\usepackage{color}
\usepackage[
colorlinks=true,
linkcolor=blue,
citecolor=blue,
urlcolor = blue]{hyperref}

\begin{document}

\preprint{APS/123-QED}

\title{Super-resolution Localization of Nitrogen Vacancy Centers in Diamond with Quantum Controlled Photoswitching}

\author{Pengfei Wang}
\email{These authors contributed equally to this work.}
\affiliation{Hefei National Laboratory for Physical Sciences at the Microscale  and Department of Modern Physics, University of Science and Technology of China , Hefei, China}
\affiliation{CAS Key Laboratory of Microscale Magnetic Resonance, University of Science and Technology of China, Hefei, China}
\affiliation{Synergetic Innovation Center of Quantum Information and Quantum Physics, University of Science and Technology of China, Hefei, China}

\author{You Huang}
\email{These authors contributed equally to this work.}
\affiliation{Hefei National Laboratory for Physical Sciences at the Microscale  and Department of Modern Physics, University of Science and Technology of China , Hefei, China}
\affiliation{CAS Key Laboratory of Microscale Magnetic Resonance, University of Science and Technology of China, Hefei, China}
\affiliation{Synergetic Innovation Center of Quantum Information and Quantum Physics, University of Science and Technology of China, Hefei, China}

\author{Maosen Guo}
\email{These authors contributed equally to this work.}
\affiliation{Hefei National Laboratory for Physical Sciences at the Microscale  and Department of Modern Physics, University of Science and Technology of China , Hefei, China}
\affiliation{CAS Key Laboratory of Microscale Magnetic Resonance, University of Science and Technology of China, Hefei, China}
\affiliation{Synergetic Innovation Center of Quantum Information and Quantum Physics, University of Science and Technology of China, Hefei, China}

\author{Mengze Shen}
\affiliation{Hefei National Laboratory for Physical Sciences at the Microscale  and Department of Modern Physics, University of Science and Technology of China , Hefei, China}
\affiliation{CAS Key Laboratory of Microscale Magnetic Resonance, University of Science and Technology of China, Hefei, China}
\affiliation{Synergetic Innovation Center of Quantum Information and Quantum Physics, University of Science and Technology of China, Hefei, China}

\author{Pei Yu}
\affiliation{Hefei National Laboratory for Physical Sciences at the Microscale  and Department of Modern Physics, University of Science and Technology of China , Hefei, China}
\affiliation{CAS Key Laboratory of Microscale Magnetic Resonance, University of Science and Technology of China, Hefei, China}
\affiliation{Synergetic Innovation Center of Quantum Information and Quantum Physics, University of Science and Technology of China, Hefei, China}

\author{Mengqi Wang}
\affiliation{Hefei National Laboratory for Physical Sciences at the Microscale  and Department of Modern Physics, University of Science and Technology of China , Hefei, China}
\affiliation{CAS Key Laboratory of Microscale Magnetic Resonance, University of Science and Technology of China, Hefei, China}
\affiliation{Synergetic Innovation Center of Quantum Information and Quantum Physics, University of Science and Technology of China, Hefei, China}

\author{Ya Wang}
\affiliation{Hefei National Laboratory for Physical Sciences at the Microscale  and Department of Modern Physics, University of Science and Technology of China , Hefei, China}
\affiliation{CAS Key Laboratory of Microscale Magnetic Resonance, University of Science and Technology of China, Hefei, China}
\affiliation{Synergetic Innovation Center of Quantum Information and Quantum Physics, University of Science and Technology of China, Hefei, China}

\author{Chang-Kui Duan}
\affiliation{Hefei National Laboratory for Physical Sciences at the Microscale  and Department of Modern Physics, University of Science and Technology of China , Hefei, China}
\affiliation{CAS Key Laboratory of Microscale Magnetic Resonance, University of Science and Technology of China, Hefei, China}
\affiliation{Synergetic Innovation Center of Quantum Information and Quantum Physics, University of Science and Technology of China, Hefei, China}
\affiliation{Department of Physics, University of Science and Technology of China, Hefei, China}

\author{Fazhan Shi}
\affiliation{Hefei National Laboratory for Physical Sciences at the Microscale  and Department of Modern Physics, University of Science and Technology of China , Hefei, China}
\affiliation{CAS Key Laboratory of Microscale Magnetic Resonance, University of Science and Technology of China, Hefei, China}
\affiliation{Synergetic Innovation Center of Quantum Information and Quantum Physics, University of Science and Technology of China, Hefei, China}

\author{Jiangfeng Du}
\email{djf@ustc.edu.cn}
\affiliation{Hefei National Laboratory for Physical Sciences at the Microscale  and Department of Modern Physics, University of Science and Technology of China , Hefei, China}
\affiliation{CAS Key Laboratory of Microscale Magnetic Resonance, University of Science and Technology of China, Hefei, China}
\affiliation{Synergetic Innovation Center of Quantum Information and Quantum Physics, University of Science and Technology of China, Hefei, China}	
\date{\today}

\begin{abstract}
We demonstrate the super-resolution localization of the nitrogen vacancy centers in diamond by a novel fluorescence photoswitching technique based on coherent quantum control. The photoswitching is realized by the quantum phase encoding based on pulsed magnetic field gradient. Then we perform super-resolution imaging and achieve a localizing accuracy better than $1.4~\mathrm{nm}$ under a scanning confocal microscope. Finally, we show that the quantum phase encoding plays a dominant role on the resolution, and a resolution of $0.15~\mathrm{nm}$ is achievable under our current experimental condition. This method can be applied in subnanometer scale addressing and control of qubits based on multiple coupled defect spins.

\end{abstract}

\maketitle



Owing to the long coherence time and the quantum scalability to multi-qubit system, solid-state defects with spins, such as the nitrogen-vacancy (NV) centers, silicon vacancy centers in diamond \cite{gruber1997scanning} and defect centers in silicon carbide \cite{koehl2011room}, become the fundamental building blocks for scalable quantum computers \cite{neumann2010quantum,dolde2013room-temperature,shi2010room} and high sensitive quantum sensors with nanoscale resolution \cite{balasubramanian2008nanoscale,maze2008nanoscale}. Localization and quantum manipulation of these defect centers at nanometer scale is of great significance for their utilization in the quantum devices \cite{scarabelli2016nanoscale,sipahigil2016integrated}, biomarker in cells \cite{mcguinness2011quantum,kucsko2013nanometre-scale} and magnetic imaging \cite{sage2013optical,devience2015nanoscale,steinert2013magnetic,simpson2017electron}. However, the conventional optically detected magnetic resonance based on confocal microscopy has a diffraction limited resolution that is not able to resolve the quantum system such as two or three coupled NV centers separated in the distance below $10~\mathrm{nm}$ \cite{neumann2010quantum}.

In the past decades, several approaches of super-resolution imaging, such as stimulated emission depletion (STED) \cite{rittweger2009sted}, ground-state depletion (GSD) \cite{han2010metastable} and stochastic optical reconstruction microscopy (STROM) \cite{pfender2014single-spin}, are implemented on the NV centers with nanometer resolution inside bulk and nanocrystal diamond. However, these optical methods either require high laser power \cite{rittweger2009sted} or the resolution is dependent on the excitation wavelength and optical intensity \cite{pfender2014single-spin}. In addition, recent advances \cite{neumann2010quantum,dolde2013room-temperature,shi2010room} employ the degree of spin resonance technique for photoswitching of neighboring NV centers as well as other defect centers. This approach enables the super-resolution localization under confocal or wide-field microscopy with much lower laser power and is independent on the optical wavelength. Nevertheless, the spatial resolutions of this approach and the above optical methods are limited, which is more than $10~\mathrm{nm}$. Moreover, the frequency encoding method exploits the different Zeeman splitting of different orientation NV centers under magnetic field \cite{chen2013wide-field} to realize the photoswitching. High DC magnetic field gradient (MFG) on the order of $10^{5}-10^{6}$ $\mathrm{T}/$$\mathrm{m}$ generated by a magnetic tip or a strong current is used for frequency encoding to achieve sub-nanometer resolution \cite{zhang2017selective,grinolds2014subnanometre}, but they need to suppress the instability or drift of the gradient magnetic field and magnetic noise. Furthermore, pushing both the resolution and the localization accuracy below $1~\mathrm{nm}$ remains challenging.

\begin{figure}
	\includegraphics[scale=1.0]{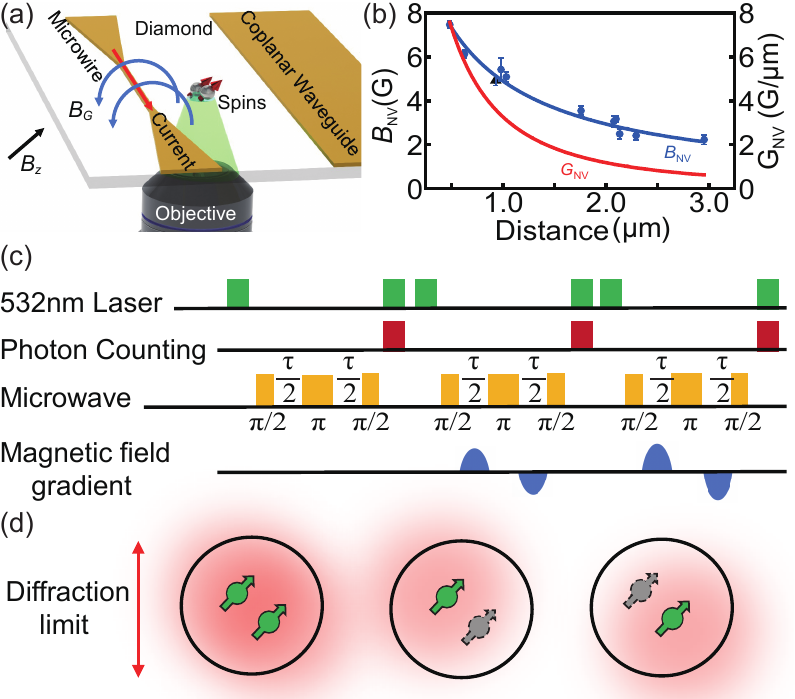}
	\caption{\label{fig:setup}Schematic view of the experimental setup and the principle of photoswitching of NV centers. (a) Experimental setup. $B_z=9.37~\mathrm{mT}$ is the static magnetic field. (b) Measured magnetic field versus the distance between the selected 11 NV centers (blue dots) and the gradient microwire by magnetic resonance spectrum and the calculated gradient (red curve) of the magnetic field projected on the NV [111] axis under a current of $I=10~\mathrm{mA}$. The magnetic field is fitted by inverse scale function and the MFG curve is the derivative of the magnetic field. (c) The spin echo pulse sequence with quantum phase encoding under pulsed MFG. (d) the schematic view of photoswitching of each NV centers corresponding to the pulse sequence in (c).}
\end{figure}

Here we adopt pulsed MFG based quantum phase encoding for the photoswitching and combine it with the confocal microscopy to investigate the localization accuracy and resolution of the NV centers. In our method, we fabricate a microwire on the diamond surface that enables the generation of MFG pulse. Then we prepare the quantum states of two neighboring NV centers within diffraction limit to fluorescence bright and dark state alternatively by the coherent quantum control of microwave and MFG pulse, respectively. The super-resolution localization is accomplished by confocal scanning and fitting the point-spread function of each NV center with the other one switching to “OFF”. Finally, we achieve an optically localization accuracy of nearby NV centers below $1.4~\mathrm{nm}$ and an ultimate photoswitching resolution of $0.15~\mathrm{nm}$ in our experimental condition. Based on pulsed MFG quantum phase encoding, the photoswitching resolution is greatly enhanced. Remarkably, our approach has the key advantages of no stringent requirement for high laser power or high MFG, and it is independent on the optical wavelength. 

The experimental setup is based on \cite{wang2015high} and the scanning head is shown in Fig.~\ref{fig:setup}(a). The NV center consists of a substitutional nitrogen atom associated with a vacancy at neighboring lattice in the diamond crystal. It is a quantum emitter that fluoresces with a zero-phonon line at $637~\mathrm{nm}$ in red under illumination of $532~\mathrm{nm}$ laser. The electronic structure of the negatively charged NV center has a spin-triplet ground state, with a zero-field splitting $D_{\mathrm{GS}}=2.87~\mathrm{GHz}$. An external magnetic field is applied to eliminate the degenerate energy level of the ground state. For simplicity, we address the two-level subsystem $|0\rangle$ and $|1\rangle$ as $m_{s}=0$ and $m_{s}=+1$ state of the NV center. By optically pumping under microseconds of $532~\mathrm{nm}$ laser pulse, the NV center is initialized to $|0\rangle$ at a probability $>95$ $\%$ \cite{dutt2007quantum}, as a fluorescence bright state as “ON”, while $|1\rangle$ acts as a fluorescence dark state as “OFF”, for the super-resolution localization applications.

The NV centers adopted here are generated by implanting $70~\mathrm{keV}$ ${}^{14}\mathrm{N}$ ions with a dose of $2\times10^{9}$ $\mathrm{cm}^{-2}$ into an ultra-pure single crystal diamond (ElementSix) followed by high temperature ($1000~{}^\circ\mathrm{C}$) and high vacuum ($10^{-6}$$\mathrm{Pa}$) annealing. The typical decoherence time of the generated NV centers is longer than $200~\upmu \mathrm{s} $. The spin state is coherently manipulated by the pulsed MW magnetic field and a one-dimensional MFG delivered separately by two gold microwires with the size of $20~\upmu \mathrm{m} $ (width) $\times$ $200~\mathrm{nm} $ (thickness) and $1~\upmu \mathrm{m} $ (width) $\times$ $200~\mathrm{nm} $ (thickness), respectively. A voltage controlled current source (Stanford CS580) connected to an arbitrary waveform generator (Rigol DG5102) is used to send current to the gradient microwire. We measure the Zeeman splitting of the NV center spins at different locations \cite{Supplementarymaterial} when sending microampere currents to the gradient microwire and then calculate the MFG projected on NV [111] axis. The maximum MFG achieves $0.735~\mathrm{mT}/$$ \upmu \mathrm{m} $ while the NV center locates at $0.5~\upmu \mathrm{m} $ away from the edge of the gradient microwire (NV A).
\begin{figure}
	\includegraphics[scale=1.0]{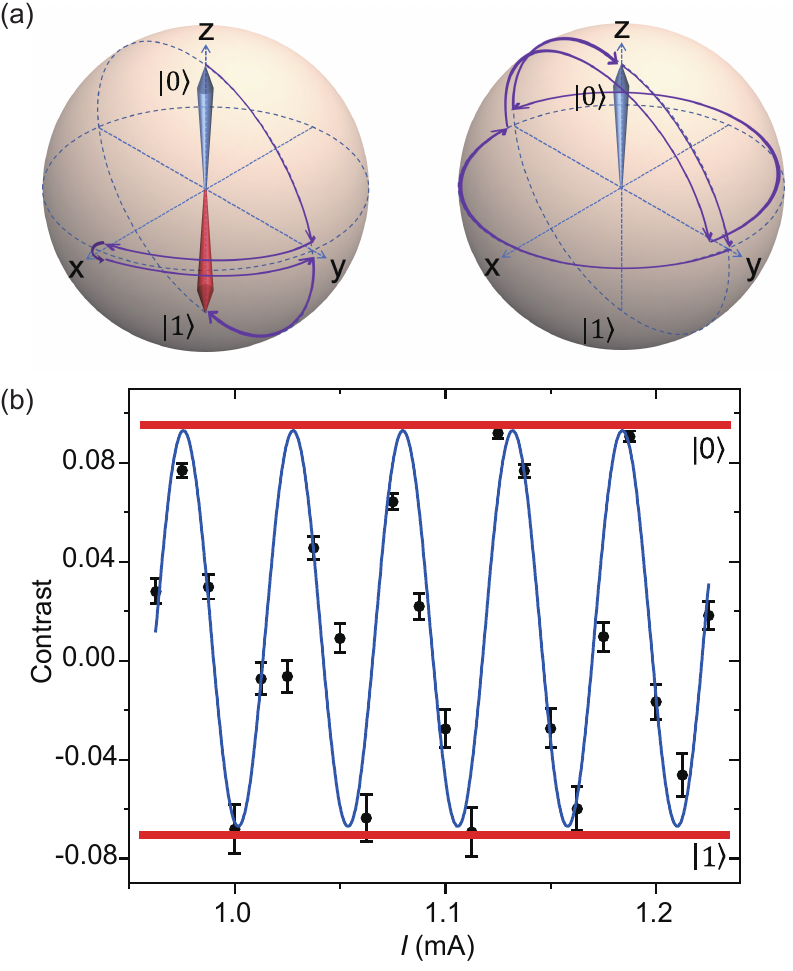}
	\caption{\label{fig:exp1} Quantum phase encoding of a single NV center spin under pulsed MFG. (a) Quantum state evolution path represented in Bloch sphere with $\varphi=\uppi$ and $\varphi=2\uppi$ and the final state of $|1\rangle$ (dark) and $|0\rangle$ (bright), respectively. (b) The normalized fluorescence intensity of NV center under $\tau=160$ $\upmu \mathrm{s} $ versus amplitude of the current sending to the microwire. The data is fitted by cosine curve.}
\end{figure}

The experiments begin with the gradient quantum control of a single NV center (NV B) with the pulse sequences shown in Fig.~\ref{fig:setup}(c). Inspired by the conventional magnetic resonance imaging (MRI) technique, we employ the pulse sequence similar to the spin echo based on phase encoding procedure to manipulate the spin state. The first $\uppi/2$ pulse prepares the spin state into quantum superposition state $(|0\rangle+|1\rangle)/{\sqrt{2}}$. Then during the evolution, there is an accumulated relative phase $\varphi_j=\gamma\tau\int_{0}^{\tau} G(r_j,t,I)r_j\, dt$ on the quantum superposition state due to the pulsed MFG, where $\gamma$ is the gyromagentic ratio, $\tau$ is the total evolution time and $G(r,t,I)$ is the pulsed MFG (Fig.~\ref{fig:exp1}). The second $\uppi/2$ pulse manipulates the quantum state into the final state $\Psi_j={\mathrm{cos}}\varphi_j|0\rangle+i{\mathrm{sin}}\varphi_j|1\rangle$. The normalized fluorescence intensity is proportional to the cosine of the relative phase: 
\begin{equation}
s_j \sim {\mathrm{cos}}(\gamma\tau\int_{0}^{\tau} G(r_j,t,I)r_j\, dt),
\end{equation}
By sending various amplitude of the cosine shaped AC current in the microwire to varying MFG, the final state is manipulated between $|0\rangle$ and $|1\rangle$ corresponding to $\varphi=2n\uppi$ and $\varphi=(2n+1)\uppi$ respectively, as shown in Fig.~\ref{fig:exp1}(b) for the case of a single NV center.

\begin{figure}
	\includegraphics[scale=1.0]{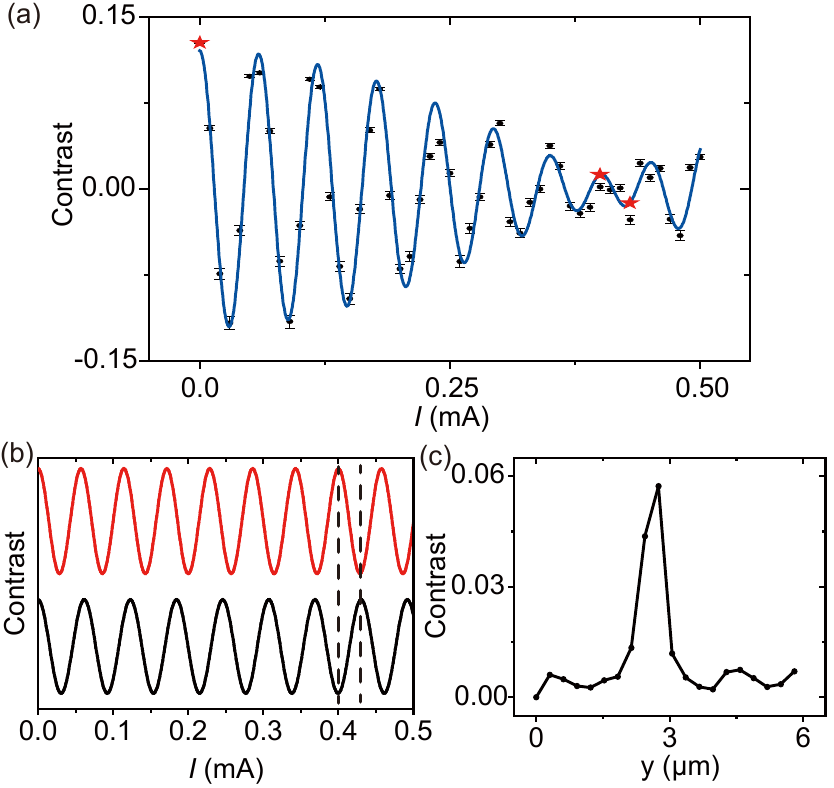}
	\caption{\label{fig:exp2} Quantum state control of NV C $\&$ D under $\tau=20$ $\upmu \mathrm{s} $. (a) Normalized fluorescence intensity of two NV centers versus the amplitude of the current in microwire. The red pentagons indicate the final state set of $|0\rangle\otimes|0\rangle$, $|0\rangle\otimes|1\rangle$ and $|1\rangle\otimes|0\rangle$. The data is fitting by cosine curve with two different frequencies. (b) Illustration of the state evolution of each NV center decomposed from (a). The dashed lines are corresponding to each quantum state in (a). (c) Real space image obtained from fast Fourier transformation of (a). The pixel resolution is estimated to be about $300~\mathrm{nm}$.}
\end{figure}

The pulsed MFG enables the spatially selective manipulations of multiple spins that are in one diffraction limited confocal fluorescence spot with same orientation and thus enables photoswitching down to nanometer scale. The NV centers at different locations accumulate position-dependent quantum phase and are able to be manipulated to different final spin states. The final spin state containing all the NV centers in one spot can be written as:
\begin{equation}
\Psi = \Psi_1\otimes\Psi_2\otimes\Psi_3\cdots,
\end{equation}
The total optical signal can be written as
\begin{equation}
s = \sum_{j}C_j{\mathrm{cos}}(\gamma\tau\int_{0}^{\tau} G(r_j,t,I) \times r_j\, dt),
\end{equation}
where $C_j$ is the normalized fluorescence intensity of each NV center.

\begin{figure}
	\includegraphics[scale=1.0]{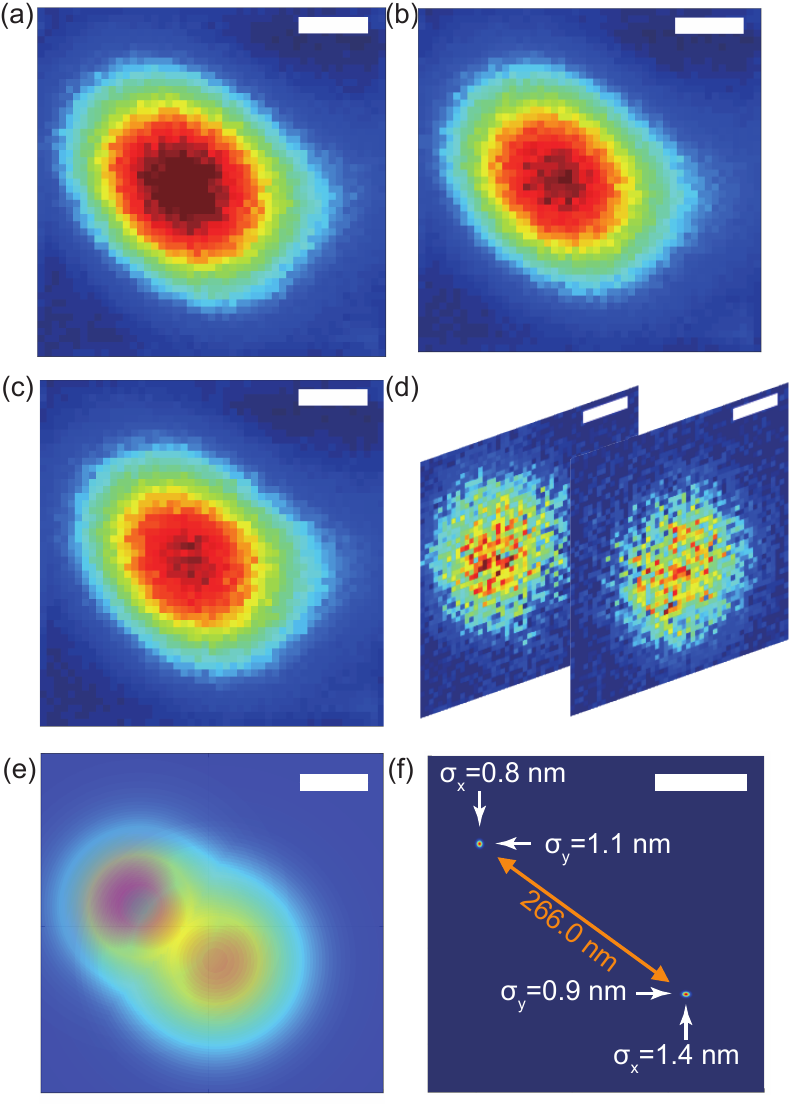}
	\caption{\label{fig:exp3} Experimental results of super-resolution localization of two neighboring NV centers. (a-c) the confocal scan results of the three sets of quantum states from left to right: $F_{|0\rangle\otimes|0\rangle}$, $F_{|0\rangle\otimes|1\rangle}$ and $F_{|1\rangle\otimes|1\rangle}$, respectively. (d) The subtraction results of the confocal scan image $F_{|0\rangle\otimes|0\rangle}-F_{|1\rangle\otimes|0\rangle}$ and $F_{|0\rangle\otimes|0\rangle}-F_{|0\rangle\otimes|1\rangle}$ reveals the position of each NV center. (e) The two-dimensional Gaussian fitting results of (b). (f) The relative two-dimensional location of the two NV centers. $\upsigma_x$, $\upsigma_y$ is the fiting error of (d) on x and y direction, respectively. Scale bars in (a-e): $200~\mathrm{nm}$. Scale bar in (f): $100~\mathrm{nm}$.}
\end{figure}

We exploit two neighboring NV centers (NV C $\&$ D) within diffraction limit to perform the spatially selective spin state manipulation. The two NV centers are $2~\upmu \mathrm{m} $ away from the edge of the microwire, making the maximum MFG estimating to be $\sim$ $0.024~\mathrm{mT}/$$ \upmu \mathrm{m} $ at $2~\mathrm{mA}$. In this case, as shown in Fig.~\ref{fig:exp2}, when varying the current sending to the microwire, the two NV centers are manipulated to $|0\rangle$ or $|1\rangle$ with a different rate due to the MFG. The fluorescence intensity signal is the sum of the two NV centers so an oscillation with a modulation envelope is observed. At the particular point that the accumulated quantum phases on two NV centers differ $n\uppi$ ($n$ is an odd number), the final spin state could be manipulated to approximate $|0\rangle\otimes|1\rangle$ or $|1\rangle\otimes|0\rangle$ at a current of $0.40~\mathrm{mA}$ or $0.43~\mathrm{mA}$, respectively (see more details for Rabi and free induction decay measurements in \cite{Supplementarymaterial}). According to the relationship between the fluorescence intensity and spin state where we define spin state $|0\rangle$ as “ON” state and $|1\rangle$ as “OFF” state, these two states can thus be switched by controlling the MFG.

We then demonstrate the super-resolution imaging on the two neighboring NV centers (Fig.~\ref{fig:exp3}). A home-build confocal microscopy is employed to scan the fluorescence spot and locate each NV center. The diffraction limited spot size defined by full width at half maximum (FWHM) is $436~\mathrm{nm}$ on the $\mathrm{NA}=0.7$ objective lens (Olympus LUCPLFLN60X) and $30~\upmu \mathrm{m} $ pinhole. In our demonstration, we consecutively take three fluorescence intensity maps for the super-resolution imaging (Fig.~\ref{fig:exp3}(a-c)). In the first image, the two NV centers are both set to “ON” state as background. When taking the last two fluorescence intensity maps, similar to STORM, one NV center at “ON” state is being imaged with the other NV center setting to “OFF” state. In order to increase the signal to noise ratio and reduce the uncertainty due to thermal drift, we apply the pulse sequence plotted in Fig.~\ref{fig:setup}(c) on these adjacent two NV centers and take an average of the whole pulse sequence for $ 2\times10^{5}$ times. The total acquistion time for the whole process is about $10h$. Finally, three images of $|0\rangle\otimes|0\rangle$, $|0\rangle\otimes|1\rangle$ and $|1\rangle\otimes|0\rangle$ states with $50\times50$ pixels and size of $1\times1$ ${\upmu\mathrm{m}}^{2}$ are obtained. The image of each single NV center is calculated by subtraction of $|0\rangle\otimes|0\rangle$ from $|0\rangle\otimes|1\rangle$ and $|1\rangle\otimes|0\rangle$. The current sending to the microwire is weak so that no thermal drift is observed during the scan. Fitting the fluorescence spot by Gaussian shaped surface with the peak value of which is the localization of the NV center, we acquire that the two NV centers at a distance of $266.0~\mathrm{nm}$ are resolved with an accuracy of $(0.8~\mathrm{nm},1.1~\mathrm{nm})$ and $(1.4~\mathrm{nm},0.9~\mathrm{nm})$. This is in excellent agreement to the result based on one-dimensional Fast Fourier Transform magnetic imaging \cite{Supplementarymaterial}.

The localization of the two NV centers separated $\sim$ $266.0~\mathrm{nm}$ does not show the ultimate capability of this technique. On the one hand, similar to the super-resolution localization microscopy, the localization accuracy is mostly dependent on the photon shot noise that can be reduced to sub nanometer by long time accumulation \cite{pfender2014single-spin}. On the other hand, the spatial resolution relies on the resolution of gradient quantum control. To distinguish the each nearby NV centers, the minimum condition is manipulating them to $|0\rangle\otimes|1\rangle$ and $|1\rangle\otimes|0\rangle$ state for photoswitching at the maximum magnetic gradient field. This is equivalent to the accumulated quantum phases $\varphi_j$ differ $\uppi$ at close to the maximum MFG. At the position of NV A with the MFG of $0.735~\mathrm{mT}/$$ \upmu \mathrm{m} $, the minimum resolution of gradient quantum control achieves $0.15~\mathrm{nm}$ \cite{Supplementarymaterial}.

In conclusion, we have demonstrated the photoswitching technique of the neighboring NV centers by a pulsed MFG quantum phase encoding and localized each NV center to $\sim$ $1.4~\mathrm{nm}$ accuracy on a confocal microscope. The minimum spatial resolution of photoswitching can achieve subnanometer and the localization accuracy can be further improved to subnanometer as well under the experimental condition shown in this work. This method provides a novel way for precisely coherent control or qubit operation of multi coupled defect spins in diamond (see more details for qubit operation of coupled NV centers in \cite{Supplementarymaterial}) and integration into low power consumption quantum chip. As it is based on the manipulation of spins, this technique could be directly applied to other solid-state spins such as silicon carbide defect spins. Combining with the wide field microscopy, it allows to build a wide-field super-resolution defect spin microscope with the capability of super-resolution photoswitching on a bulk diamond sensor for magnetic imaging and biomarker imaging.

The authors acknowledge Y. H. Lin and K. W. Xia for helpful discussions. This work was partially carried out at the USTC Center for Micro and Nanoscale Research and Fabrication. This work was supported by the National Key Research and Development Program of China (grant no. 2018YFA0306600, 2018YFF01012500, 2016YFA0502400), the National Natural Science Foundation of China (grant nos. 11874338, 81788101, 91636217, 11722544, 11761131011, and 31971156), the CAS (grant nos. GJJSTD20170001, QYZDY-SSW-SLH004 and YIPA2015370), the Anhui Initiative in Quantum Information Technologies (grant no. AHY050000), the Anhui Provincial Natural Science Foundation (grant nos. 1808085J09), the CEBioM, the national youth talent support program, the Fundamental Research Funds for the Central Universities.

%


\end{document}